\documentclass{article}
\usepackage[preprint]{spconf}
\usepackage{amsmath,graphicx}
\usepackage{shortcuts}
\usepackage{svg}
\usepackage{optidef}
\usepackage{booktabs}
\usepackage{caption}
\usepackage{subcaption}
\usepackage{url}

\newcommand{\nsynth}[1]{}

\makeatletter
\def\blfootnote{\gdef\@thefnmark{}\@footnotetext}
\makeatother

\title{Unsupervised Harmonic Parameter Estimation Using Differentiable DSP and Spectral Optimal Transport}
\name{Bernardo Torres \qquad Geoffroy Peeters \qquad Gaël Richard}
\address{LTCI, Telecom Paris, Institut Polytechnique de Paris}
\copyrightnotice{\copyright 2024 IEEE. Personal use of this material is permitted. Permission from IEEE must be obtained for all other uses, in any current or future media, including reprinting/republishing this material for advertising or promotional purposes, creating new collective works, for resale or redistribution to servers or lists, or reuse of any copyrighted component of this work in other works.}

\begin{document}
\ninept
\maketitle
\begin{abstract}

In neural audio signal processing, pitch conditioning has been used to enhance the performance of synthesizers. However, jointly training pitch estimators and synthesizers is a challenge when using standard audio-to-audio reconstruction loss, leading to reliance on external pitch trackers. To address this issue, we propose using a spectral loss function inspired by optimal transportation theory that minimizes the displacement of spectral energy. We validate this approach through an unsupervised autoencoding task that fits a harmonic template to harmonic signals. We jointly estimate the fundamental frequency and amplitudes of harmonics using a lightweight encoder and reconstruct the signals using a differentiable harmonic synthesizer. The proposed approach offers a promising direction for improving unsupervised parameter estimation in neural audio applications.

\end{abstract}

\blfootnote{This work was funded by the European Union (ERC, HI-Audio, 101052978). Views and opinions expressed are however those of the author(s) only and do not necessarily reflect those of the European Union or the European Research Council. Neither the European Union nor the granting authority can be held responsible for them.}
\section{Introduction}

\label{sec:intro}
Spectral Modeling Synthesis (SMS) \cite{serraSpectralModelingSynthesis1990}, a foundation in signal processing, aims to represent audio signals through a combination of sinusoidal and noise components that define clear spectral structures visible in short-term spectral analysis. A significant portion of work on SMS has been dedicated to modeling harmonically related sinusoidal signals, where a fundamental frequency parameterizes a harmonic template \cite{mcauleyspeech}. The task is threefold: identifying the fundamental frequency, estimating harmonic amplitudes, and capturing transients and aperiodicities. While existing methods offer suitable solutions, they often fall short in dynamic scenarios where frequency amplitudes vary over time or in data-driven systems trained end-to-end via gradient descent.

Recent advances in Differentiable Digital Signal Processing (DDSP) \cite{engelDDSPDifferentiableDigital2020} have sought to bridge traditional signal processing with the capabilities of deep learning. These methods often depend on an external pitch estimator for determining the fundamental frequency, subsequently used for conditioning synthesizers \cite{hayesReviewDifferentiableDigital2023}. This approach favors harmonic alignment but constrains their flexibility and adaptability. On the other hand, Deep Neural Networks have outperformed traditional DSP-based methods in pitch estimation but still, for the most part, rely on supervised learning. Additionally, they might suffer from poor generalization due to data distribution mismatches \cite{morrisonCrossdomainNeuralPitch2023}, and poor integration with other tasks.

Some noteworthy approaches, such as SPICE \cite{gfellerSPICESelfsupervisedPitch2020} have emerged that tackle pitch estimation through analysis-by-synthesis and self-supervised learning. However, most existing methods focus either on pitch or amplitude estimation, rarely both. The seminal DDSP paper \cite{engelDDSPDifferentiableDigital2020} and subsequent works \cite{dayiwuwenyihsiaoDDSPbasedSingingVocoders2022} have shown that unsupervised approaches to joint estimation is particularly challenging, and notable recent works address this issue directly through pre-training on synthetic data \cite{engelSelfsupervisedPitchDetection2020}, or using a damped sinusoidal model \cite{hayesSinusoidalFrequencyEstimation2023}.

Recent literature has begun to question the efficacy of the commonly used audio-to-audio reconstruction losses \cite{turianSorryYourLoss2020}. Our work grows out of this branch and turns towards the theory of optimal mass transport for a novel perspective for comparing spectral representations.

Optimal transport has been employed in various audio tasks, such as Non-negative Matrix Factorization (NMF) based transcription \cite{flamaryOptimalSpectralTransportation2016}, Blind-Source separation \cite{roletBlindSourceSeparation2018}, estimating the pitch of inharmonic signals \cite{elvanderUsingOptimalTransport2017}, performing spectral interpolation \cite{hendersonAudioTransportGeneralized2019} and defining distances between power spectral densities (PSDs) of time series \cite{cazellesWassersteinFourierDistanceStationary2020}. To the best of our knowledge, this is the first work to propose an optimal transport-based cost for differentiable spectral comparison and empirical risk minimization in a neural audio system. 

This paper is organized as follows.  Firstly, Section \ref{sec:ddsp_mss}  delves into the limitations of DDSP models and \textit{vertical} spectral losses. Section \ref{sec:ot} introduces our alternative loss function, followed by a description of our proposed autoencoding task and experiments (Section \ref{sec:method}) and results (Section \ref{sec:results}). Section \ref{sec:conclusion} provides some concluding remarks.

\section{DDSP and the Multi-Scale Spectral loss}\label{sec:ddsp_mss}

DDSP synthesizers allow fine-grained spectral modeling \cite{engelDDSPDifferentiableDigital2020}. When provided with accurate fundamental frequency ($f_0$), these models have a high degree of harmonic alignment by design. DDSP-inspired works have capitalized on this property by using $f_0$ conditioning from external pitch trackers \cite{hayesReviewDifferentiableDigital2023} and optimizing spectral losses. 

 In time-frequency analysis, larger window sizes offer finer frequency discrimination but at the cost of temporal resolution. Conversely, smaller window sizes are prone to spectral leakage, although they have better time resolution. By using multiple resolutions of the Short-Time Fourier Transform (STFT), the Multi-Scale Spectral loss (MSS) hopes to achieve the best of both worlds \cite{engelDDSPDifferentiableDigital2020, wangNeuralSourcefilterWaveform2019}:

\begin{equation}
    \mathcal{L}_{\text{MSS}}(s, \hat s)=\sum_{\gamma \in \Gamma} \mathcal{L}_{\text{lin}}^\gamma + \mathcal{L}_{\text{log}}^\gamma,
\end{equation}\label{eq:msstft}

\vspace{-3pt}

\noindent where ($s$, $\hat s$) are time domain signals. The linear and log spectral losses are $\mathcal{L}_{\text{lin}}^\gamma = \left\|\mathbf{S}_\gamma - \mathbf{\hat S}_\gamma\right\|_{1}$  and $\mathcal{L}_{\text{log}}^\gamma =  \left\|\log (\mathbf{S}_\gamma)- \log(\mathbf{\hat S}_\gamma)\right\|_{1}$, respectively, where $\mathbf{S}_\gamma = \left|\text{STFT}_{\gamma}(s) \right|$ is the magnitude STFT at scale (window size) $\gamma$. The set of scales $\Gamma$ usually consists of powers of 2. The linear loss is more effective during the initial training stages \cite{arikFastSpectrogramInversion2018} by emphasizing large amplitude differences, while the logarithmic loss may help in discerning more subtle frequency patterns.


Following the terminology in \cite{cazellesWassersteinFourierDistanceStationary2020}, we refer to losses that operate point-wise between spectra with identical support as \textit{vertical} losses. Conventional spectral losses fall under this category, comparing using an $L_p$ norm the amplitude of a given frequency bin of a reference frame spectrum to the amplitude of the same bin on a target. If the spectra do not overlap, a \textit{vertical} comparison may not properly measure the amount of energy displacement on the frequency axis.


When attempting to learn a pitch tracker with the neural audio model in an end-to-end fashion, the MSS (a \textit{vertical} loss) has been reported as ineffective \cite{engelDDSPDifferentiableDigital2020,dayiwuwenyihsiaoDDSPbasedSingingVocoders2022}, which may be due to its poor sense of gradient orientation with respect to an oscillator's frequency \cite{turianSorryYourLoss2020, masudaImprovingSemiSupervisedDifferentiable2023} and many notable local minima \cite{hayesSinusoidalFrequencyEstimation2023}.

 A \textit{horizontal} comparison of spectra should faithfully represent spectral energy displacement. In this context, we aim to validate an alternative loss function that enables a more nuanced, \textit{horizontal} comparison of spectral content, thereby guiding neural audio models toward more accurate frequency estimation. The framework of Optimal Transport offers precisely that.

\section{Optimal transport}\label{sec:ot}


Let $\alpha = \sum^n_{i=1} \mathbf{a}_i \delta_{x_i}$ and $\beta = \sum^m_{j=1} \mathbf{b}_j \delta_{y_j}$ be discrete measures with weights $\mathbf{a}_i$ at positions $x_i$ and weights $\mathbf{b}_j$  at positions $y_j$, where $\delta_{x}$ is the Dirac measure at location $x$. 
Let additionally ($\mathbf{a}$, $\mathbf{b}$)  belong to the space of probability vectors, $i.e.$ $\mathbf{a} \in \Sigma_n$ and $\mathbf{b} \in \Sigma_m$, for $\Sigma_n = \left\{\mathbf{a} \in \mathbb{R}^n_+ ; \sum_{i=1}^n \mathbf{a}_i = 1 \right\}$. Given a cost function $c(x_i, y_j)$ representing the cost of transporting a unit of mass from $x_i$ to $y_j$, the discrete Optimal Transport (OT) problem is  \cite{peyreComputationalOptimalTransport2019}:



\begin{equation}
    \mathcal{L}_c(\alpha, \beta) = \min_{\mathbf{P}}  \sum_{i=1}^n \sum_{j=1}^m \mathbf{P}_{ij} c(x_i, y_j),
    \label{eq:kontorovitch}
\end{equation}

\noindent where $\mathbf{P} \in \mathbb{R}^{n \times m}_+$ is the transport plan describing the amount of mass to be transported from bin $i$ to bin $j$, subject to mass conservation constraints $ \{\mathbf{a}_i = \sum_{j=1}^m \mathbf{P}_{ij}\}_{i=1,..,n}$ and $\{ \mathbf{b}_j = \sum_{i=1}^n \mathbf{P}_{ij} \}_{j=1,..,m}$. The optimal transport map, $\mathbf{P}^*$, gives the optimal way to transport $\alpha$ to $\beta$. Note that $\mathcal{L}_c(\alpha, \beta)$ depends on both the weights ($\mathbf{a}$, $\mathbf{b}$), and the supporting positions ($x_i$, $y_j$) \cite{peyreComputationalOptimalTransport2019}.

For cost functions $c(x_i, y_j) = \left| x_i - y_j \right|^p$, $\mathcal{L}_c(\alpha, \beta)^{1/p}$ is the p-Wasserstein distance  between $\alpha$ and $\beta$ \cite{kolouriOptimalMassTransport2017}. This distance allows for the comparison of distributions when they do not overlap, which is a limitation of $L_p$ norm-based distances and commonly employed divergences (such as the Kullback–Leibler).

\subsection{Optimal transport in one dimension}

\begin{figure}
    \centering
    \hfill
    \begin{subfigure}[b]{0.93\columnwidth}
        \centering
        \includegraphics[width=\textwidth]{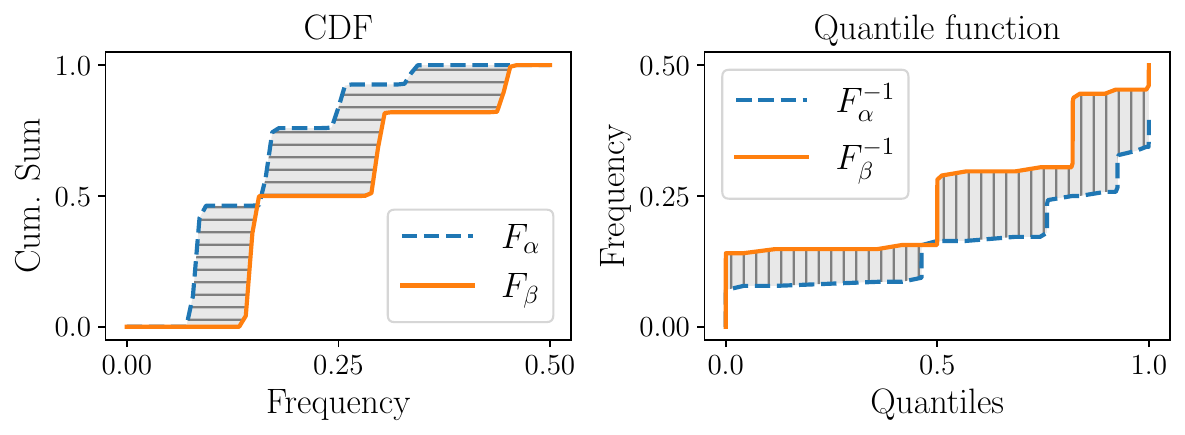}
    \end{subfigure}
    \begin{subfigure}[b]{0.85\columnwidth}
        \centering
        \includegraphics[width=\textwidth]{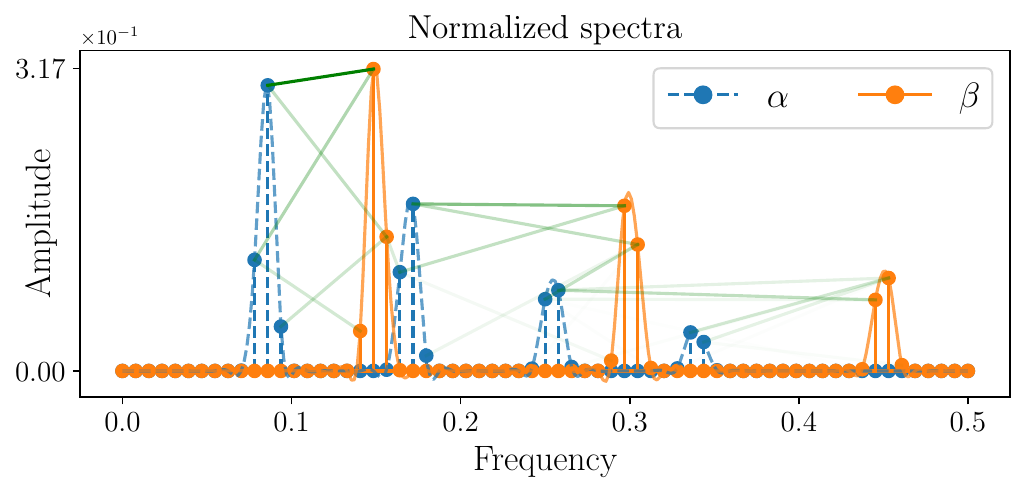}
    \end{subfigure}
    \captionsetup{belowskip=-20pt}
    \caption{Cumulative sum and quantile functions of two measures $\alpha$ and $\beta$ representing harmonic spectra with different $f_0$s across the Nyquist range [$0$-$0.5$]. The grey lines depict the pointwise differences between the inverse CDFs as per Eq. \ref{eq:wasserstein_discrete}, with the shaded region representing the Wasserstein metric $\mathcal{W}_1$.  Green connectors denote the optimal plan ($\mathbf{P}^*_{ij}$), weighted by their respective magnitudes. }
    \label{fig:cdfs_example}
\end{figure}

  While being challenging for dimensions $d>1$, the OT problem in 1D  for \emph{general} measures $(\alpha, \beta) \in \mathcal{M(\mathcal{X})}$ (the set of Radon measures on the space $\mathcal{X}$) has a closed-form solution for the p-Wasserstein distance \cite{peyreComputationalOptimalTransport2019, santambrogioOptimalTransportApplied2015}:

\begin{equation}
    \mathcal{W}_p(\alpha, \beta)^{p} =  \int_0^1 \left| F^{-1}_{\alpha}(r) - F^{-1}_{\beta}(r) \right|^p dr ,
    \label{eq:1d_ot}
\end{equation}

\noindent where $F_\alpha$ : $\mathbb{R}$ $ \to $ $[0, 1]$ is the cumulative distribution function (CDF)  $F_{\alpha}(x) = \int_{-\infty}^x d\alpha $ and $F^{-1}_\alpha$ : $[0, 1]$ $\to$ $\mathbb{R}$ is its pseudoinverse: 

\begin{equation}
    F^{-1}_{\alpha}(r) = \inf \left\{ x \in \mathbb{R} : F_{\alpha}(x) \geq r \right\},
\end{equation}

 \noindent the generalized quantile function. When dealing with discrete measures, Equation \ref{eq:1d_ot} can be approximated by a discrete sum \cite{flamaryPOTPythonOptimal2021}:

\begin{equation}
    \mathcal{W}_p(\alpha, \beta)^{p} =  \sum_{i=1}^n \left| F^{-1}_{\alpha}(r_i) - F^{-1}_{\beta}(r_i) \right|^p (r_i - r_{i-1}), 
    \label{eq:wasserstein_discrete}
\end{equation}

\noindent where $F$ is computed using the step function $u(\cdot)$: $ F_{\alpha}(x) = \sum_{i=1}^n u(x - x_i) \mathbf{a}_i $. The set $r$ can be taken to be the ordered set of quantiles of both $\alpha$ and $\beta$ \cite{flamaryPOTPythonOptimal2021}.  Figure \ref{fig:cdfs_example} illustrates this process.

\subsection{Spectral optimal transport}

    The Wasserstein-Fourier distance was defined by Cazelles et al. \cite{cazellesWassersteinFourierDistanceStationary2020} as the $\mathcal{W}_2$ distance between Power Spectral Densities of stationary time-series signals. We refer hereafter to Spectral Optimal Transport (SOT) as a more general sense of \textit{horizontal} (measuring energy displacement) comparison of discrete time-frequency spectra by approximating $\mathcal{W}_p$ using Equation \ref{eq:wasserstein_discrete}. 
    
     Let $\mathbf{\Phi}$ $:$ $\RR^N $ $\to $ $ \RR^{L \times M}_+$ be a normalized time-frequency transform so that each frame $\mathbf{\Phi}(s)_j$ sums to 1. We define the \textit{horizontal} SOT loss $\mathcal{L}_\text{SOT}$ as the temporal mean (over $L$ time frames) of the p-Wasserstein distance computed on the frequency axis of $\mathbf{\Phi}$:

    \begin{equation}
        \mathcal{L}_\text{SOT}(s, \hat s) = \frac{1}{L} \sum_{j=0}^{L-1} \mathcal{W}_p(\mathbf{\Phi}(s)_j, \mathbf{\Phi}(\hat s)_j)^{p},
        \label{eq:sot}
    \end{equation}

\noindent where $s$ and $\hat s$ are time-domain signals. Recall that $\mathcal{W}_p$ depends on the weights ($\mathbf{a}$, $\mathbf{b}$) and their positions ($\{x_i\}_{i=1,.., M}, \{y_j\}_{j=1, .., M}$), which here represent respectively the amplitude vectors of the frame spectra and the frequencies of bins $(i, j)$ of $\mathbf{\Phi}$.

\begin{figure}[h]
  \centering
  \includegraphics[width=\columnwidth]{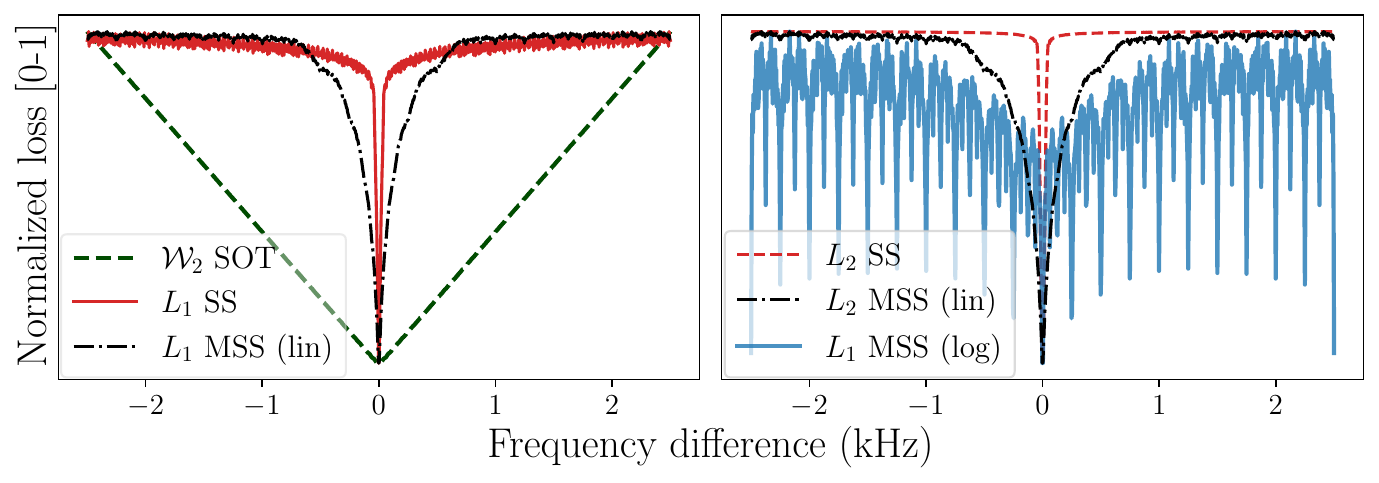}
  \captionsetup{belowskip=-0.5pt}
  \caption{Normalized \textit{spectral}  Single-Scale (SS), Multi-Scale (MSS) and proposed SOT $\mathcal{W}_2$ losses as a function of the frequency shift between two $16$ KHz sinusoidal signals of $4096$ samples. The reference sinusoid has a frequency of $4000$ Hz. SS is computed for window size $\gamma=1024$ and MSS with scales $\Gamma=\{2^k\}_{k=6,\dots ,11}$.} 
  
  \label{fig:sweep}
\end{figure}

\textbf{SOT for simple sinusoids}:   As demonstrated in \cite{cazellesWassersteinFourierDistanceStationary2020}, the Wasserstein-Fourier distance of two frequency-shifted signals is proportional to the frequency shift. We illustrate this in Figure \ref{fig:sweep}, where we show the value of different spectral losses between two sinusoidal signals as a function of their frequency difference. We compute the losses in the time-frequency domain and average over time, using the SOT formulation in Eq. \ref{eq:sot} with $\mathcal{W}_2$. SOT points towards a minimal frequency difference, while the \textit{vertical} $L_1$ and $L_2$ losses converge smoothly only for small differences. This serves as a motivating example for studying the SOT loss for more challenging spectra.

\section{Methodology and experiments}\label{sec:method}




Our approach employs an encoder for joint frame-wise estimation of fundamental frequency and harmonic amplitudes of a signal, taking as input the Constant-Q Transform (CQT). The signal is reconstructed in the time domain using a DDSP harmonic synthesizer \cite{engelDDSPDifferentiableDigital2020}, and trained end-to-end via an audio reconstruction loss. The underlying assumption of this approach is that the instantaneous frequency varies slowly enough for accurate estimation from spectral frames (local stationarity). Through this toy autoencoding task,  we study the interplay between \textit{vertical} losses (such as the MSS) and \textit{horizontal} SOT variants under various conditions.

\vspace{-3pt}

\subsection{Analysis/synthesis autoencoder}


\textbf{Input representation}: We compute the CQT from the input waveform on the fly \cite{cheukNnAudioOntheFlyGPU2020}. The CQT's logarithmic scaling along frequencies naturally relates pitch shifting in semitones as frequency-axis translations.  The lower frequency bound of the CQT is set to $32.7$ Hz (the C1 note) and $285$ bins are used to cover the whole 8 kHz spectrum (amounting to 3 bins per semitone).  A hop size of $256$ is used, which defines the estimation rate of the system.

\textbf{Encoder}: 
Our encoder adopts the lightweight architecture from \cite{riouPESTOPitchEstimation2023} ($46 K$ parameters), featuring 1D convolutions along the frequency axis. A Toeplitz linear layer\footnote{We point the reader to \cite{riouPESTOPitchEstimation2023} for a detailed explanation of the architecture} maps the output feature map from the encoder to $285$ pitch logits, which are mapped to a categorical distribution over pitch classes $\mathbf{p} \in \RR^{285}$ through a softmax operation \cite{engelSelfsupervisedPitchDetection2020}. The employed architecture ensures translation invariance, meaning that a pitch shift in the input (CQT)  corresponds to a shift in the output pitch probabilities by an equal amount of bins. 

 We set the number of estimated harmonics $H$ to $20$. Harmonic amplitudes are obtained through a feed-forward linear layer, which maps the the output feature map from the encoder to $H$ amplitude values $\mathbf{c}$, followed by an exponentiated sigmoid activation from DDSP \cite{engelDDSPDifferentiableDigital2020}. Unlike \cite{engelDDSPDifferentiableDigital2020}, we treat each spectral frame as independent. 
 
\textbf{Pitch regression:} The pitch is computed as the expectation of the pitch distribution, or the \emph{soft-argmax}: $\hat{f}^\text{u} = \sum_{i=1}^{285} f_i^\text{u} \mathbf{p}_i$; where $f_i^\text{u}$ is the frequency of the $i$-th pitch class, logarithmically mapped to the $[0,1]$ range. Each bin $f_i^\text{u}$ corresponds to a frequency bin in the input CQT. A temperature parameter $\tau=0.1$ is used to pre-scale the logits to facilitate an unimodal distribution \cite{kendallEndtoendLearningGeometry2017}.

\textbf{Harmonic synthesis:} We use a DDSP harmonic synthesizer \cite{engelDDSPDifferentiableDigital2020}: 

\begin{equation}
    s(n)=\sum_{h=0}^{H-1} \mathbf{c}_h (n) \sin \left(\phi_{h}(n)\right),
 \label{eq:harmonic}
\end{equation}

 \noindent parameterized at the sample-level by $f_0$ and amplitudes $\mathbf{c}$. The cumulative sum  $\phi_{h}(n)=2 \pi \sum_{t=0}^{n}f_h (t) $ computes the phase at time  $n$, for $f_h$ = $h f_0$. We set initial phases to $0$ and use the same frame rate to sample rate interpolation of $f_0$ and $\mathbf{c}$ as in DDSP.

\begin{table*}[]
\centering
\captionsetup{belowskip=-10pt}
\setlength{\tabcolsep}{4.8pt} 
\begin{tabular}{lllllllllllll}
\toprule
& \multicolumn{4}{c}{Variations} & \multicolumn{4}{c}{Mean (STD)} & \multicolumn{4}{c}{Median} \\
\cmidrule(lr){2-5} 
\cmidrule(lr){6-9} 
\cmidrule(lr){10-13}
& $\gamma$ ({\scriptsize$\mathcal{L}_\text{SOT}$}) & $\Gamma$ ({\scriptsize$\mathcal{L}_\text{Lin}$}) & LogF & $f_\text{cut}$ & LSD [dB] $\downarrow$ & RPA [\%] $\uparrow$& RCA [\%] $\uparrow$ & \multicolumn{1}{c}{OD} & LSD & RPA & RCA & \multicolumn{1}{c}{OD} \\ 
\midrule
\texttt{MSS-Lin} & - & $\Gamma_0$ & - & - & 46.4 (21.4) & 20.2 (44.6) & 26.9 (42.7) & -2.3 (1.3) & 58.0 & 0.2 & 3.9 & -2.8 \\
\texttt{MSS-LogLin} & - & $\Gamma_0$ & - & - & 80.5 (15.1) & 1.4 (2.7) & 4.0 (4.5) & -0.9 (1.9) & 82.6 & 0.1 & 3.2 & -0.6\\ 
\midrule
\texttt{SOT-2048} & 2048 & $\Gamma_0$ & $\times$ & $\checkmark$ & \textbf{23.5 (3.5)} & \textbf{75.0 (43.2)} & \textbf{99.2 (1.6)} & -0.5 (0.9) & \textbf{24.5} & \textbf{99.7} & \textbf{99.8} & 0.0 \\
\texttt{SOT-512} & 512 & $\Gamma_0$ & $\times$ & $\checkmark$ & 40.5 (23.5) & 42.9 (39.4) & 62.3 (42.6) & -0.8 (0.7) & 26.6 & 63.6 & 75.2 & -0.4\\
\texttt{SOT-512-LogF} & 512 & $\Gamma_0$ & $\checkmark$ & $\checkmark$ & \underline{25.9 (2.5)} & \underline{55.4 (36.1)} & \underline{86.8 (16.2)} & -0.7 (0.9) & \underline{25.0} & \underline{63.7} & \underline{95.6} & -0.4\\
\texttt{SOT-NoCut} & 2048 & $\Gamma_0$ & $\times$ & $\times$ & 70.6 (31.8) & 23.7 (30.3) & 46.0 (36.4) & -0.4 (0.4) & 77.6 & 20.0 & 45.0 & -0.2 \\
\texttt{SOT-2048-SS} & 2048 & $\{512\}$ & $\times$ & $\checkmark$ & 97.9 (32.5) & 14.1 (25.5) & 28.6 (32.6) & 0.0 (1.0) & 101.1 & 4.7 & 11.6 & 0.3 \\
\bottomrule
\end{tabular}
\caption{Experimental results for the joint $f_0$ and harmonic amplitude estimation task on synthetic data. We report the mean, standard deviation and median test metrics for 5 separate training runs. Best results are indicated in bold and second-best are underlined.}\label{tab:synthetic_results}
\end{table*}

\subsection{Data}
 We generate a synthetic dataset of $4000$ examples, where each example is a $16$ kHz signal with $N=4096$ samples. Harmonic synthesis (Eq. \ref{eq:harmonic}) is performed with a fixed instantaneous frequency during the signal length. The number of harmonics of each signal is selected randomly from [$1$-$8$]. The $f_0$ values are randomly drawn from the range [$40$-$1950$] Hz and the harmonic amplitudes from [$0.4$-$1$].  The dataset is split into $70$\% training, $20$\% validation, and $10$\% testing.

\subsection{Baselines}

The baselines are trained with the MSS loss only, for scales $\Gamma_0$=$\{2048,1024,512,256,128,64\}$ and hop size respecting $75$\% overlap. \texttt{MSS-Lin} uses only $\mathcal{L}_{\text{lin}}$ and \texttt{MSS-LogLin} uses $\mathcal{L}_{\text{log}}$+$\mathcal{L}_{\text{lin}}$. 

\subsection{Spectral Optimal Transport experiments}

We use $\mathcal{W}_2$ in $\mathcal{L}_\text{SOT}$ for computing the Wasserstein distances between frame spectra, employing the squared magnitude STFT with a \emph{flattop}\footnote{Using a window with less prominent side lobes helped with stability.} window of size $\gamma$ and hop size of $256$ as the time-frequency transform $\mathbf{\Phi}$.  In all SOT experiments, MSS was also included as a \textit{vertical} loss (only $\mathcal{L}_{\text{lin}}$, with scales $\Gamma = \Gamma_0$) to penalize $\mathcal{W}_2$ minima with low spectral overlap. This can happen, for instance, when the number of active harmonics is overestimated. This also addresses instability issues observed with the 1D Wasserstein loss alone \cite{martinezClosedformGradient1D2016}, possibly due to the normalization of the spectra. The final objective is $\mathcal{L}_\text{SOT} + \lambda \mathcal{L}_{\text{lin}}$, with $\lambda = 0.05$. We additionally employ a frame-wise frequency cutoff in the spectrum of the estimated signal, limiting energy to lower frequencies, which helped to significantly reduce octave errors.
 
\textbf{Frequency cutoff:} The process starts with proportional normalization of spectra: the target frame spectrum $\mathbf{\Phi}(s)_j$ sums to $1$, while the reconstructed frame $\mathbf{\Phi}(\hat s)_j$ is normalized to a sum potentially exceeding 1 if its energy is greater than the original. For frame $j$, $f_\text{cut}$ is determined to be the frequency where the cumulative sum of the reconstructed spectrum matches the total normalized energy of the reference\footnote{In practical terms, rather than applying a filtering operation, we ignore indices ($i$) for quantile ($r$) values over one in the computation of Equation \ref{eq:wasserstein_discrete}.} ($F_{\mathbf{\Phi}(\hat s)_j}(f_\text{cut}) =1$).

\textbf{SOT Variants:} We experiment with three main $\mathcal{L}_\text{SOT}$ variants, varying window size $\gamma$ and frequency scaling LogF: \begin{enumerate*}[label=(\Roman*)]
    \item \texttt{SOT-2048} uses $\gamma=2048$;
    
    \item \texttt{SOT-512} uses $\gamma=512$;
    \item \texttt{SOT-512-LogF} is similar to \texttt{SOT-512} but additionally scales frequencies logarithmically before computing $\mathcal{W}_2$ (\emph{i.e.} $x_i \to \log(x_i), y_j \to \log(y_j)$), thereby reducing the frequency ground cost for higher frequencies.
\end{enumerate*}




 \textbf{Ablations:}  We provide two ablations to \texttt{SOT-2048} .  First, we examined the special case \texttt{SOT-2048-SS} using a single scale spectral loss (resolution $\Gamma=\{512\}$ for $\mathcal{L}_{\text{lin}}$), and   $\lambda=0.1$. Lastly, the frequency cutoff is removed in configuration \texttt{SOT-NoCut}.
 


\subsection{Training details}
We employed the Adam optimizer with a learning rate of $1$$\times$$10^{-4}$ and a batch size of $64$ for $25k$ steps. To ensure robustness against weight initialization, each model was trained five times using different seeds for the pseudo-random number generator. Code and training recipes are available online\footnote{https://github.com/bernardo-torres/1d-spectral-optimal-transport}.

\subsection{Evaluation}

Evaluation is done using a combination of pitch estimation and signal reconstruction metrics. We employ Raw Pitch Accuracy (RPA), Raw Chroma Accuracy (RCA) \cite{polinerMelodyTranscriptionMusic2007, raffelMIR_EVALTransparentImplementation2014}, and Log Spectral Distance (LSD). RPA measures the percentage of frames with estimated pitch close to the ground truth within $0.5$ semitones. RCA is similar but allows octave errors.  The LSD is computed as:
\vspace{-5pt}

\begin{equation}
\text{LSD}(s, \hat{s}) = \frac{1}{LM} \norm{\log \left(\mathbf{S}_\gamma \right) - \log (\mathbf{\hat S}_\gamma ) }_F^2,
\end{equation}

\noindent where $\norm{\cdot}_F$ denotes the Frobenius norm of a matrix, $\mathbf{S}_\gamma = |\text{STFT}_{\gamma}(s)|$ $\in \mathbb{R}^{L \times M}$ denotes the magnitude STFT of $s$ for window size $\gamma$, with $L$ frames and $M$ frequency bins. We use $\gamma=1024$ and a hop size of $256$. The Mean Octave Difference (OD) is also included to indicate, on average, the mean deviation (in octaves) of the estimated pitch from the ground truth, highlighting the nature of the octave errors.  For each run, the checkpoint with the lowest validation reconstruction score was selected for testing. We report the mean metrics on the test set.

\section{Results}
\label{sec:results}

Table \ref{tab:synthetic_results} summarizes the experimental results for our trained models on the synthetic dataset. The high standard deviation (STD) reveals a significant dependence on initialization. The initial pitch probability distribution, determined by the neural networks' weight initialization, was a key factor impacting the model's success. 

This sensitivity to initialization is particularly pronounced in models trained only with the Multi-Scale Spectral loss (\texttt{MSS-Lin} and \texttt{MSS-LogLin}). Out of five runs, only one achieved optimal LSD, RPA, and RCA. This variability is evidenced by the large standard deviation and low median values. Interestingly, logarithmic compression (\texttt{MSS-LogLin}) resulted in worse performance.

In contrast, models using the Spectral Optimal Transport (SOT) loss demonstrated greater robustness, converging to satisfactory solutions on average. Notably, the model employing a larger window size (\texttt{SOT-2048}) outperformed all others, with a mean of  $23.5$ dB LSD, $75$\% RPA and $99.2$\% RCA.

Frequency scaling and frequency cutoffs helped performance. Logarithmic frequency scaling (\texttt{SOT-512-LogF}) led to an improvement over \texttt{SOT-512}. Logarithmic scaling makes the SOT loss less punitive for high-frequency deviations, aligning with our focus on estimating the fundamental frequency. The use of the frequency cutoff was found to be beneficial when the model overestimates the signal's energy or is initialized with many active harmonics. This prevents the gradient from skewing towards an incorrect $f_0$ after the normalization step needed for SOT loss.

 The \texttt{SOT-2048-SS} model exhibited suboptimal results in LSD, RPC, and RCA, but maintained an average octave difference of zero, indicating that the pitch solution it converged to was within the octave range of the ground truth. This is a nuanced failure mode in which the model retains more or less harmonics than actually present in the reference signal. Obtaining perfect spectral alignment by minimizing $\mathcal{W}_2$ in $\mathcal{L}_\text{SOT}$ requires matching the number of harmonics with the reference, implying that a mismatch may lead to inaccurate $f_0$ predictions. The introduction of a \textit{vertical} loss was intended to penalize solutions with low spectral alignment, but using only a single scale has proven to be insufficient.


While promising, our results also raise questions on the interplay between MSS and SOT losses. Preliminary findings suggest that these loss functions may be optimizing different aspects of the estimator and could be working in a conflicting manner. For instance, the \textit{horizontal} loss might favor frequency shifts in the spectra, while the \textit{vertical} loss may lower the amplitudes, possibly causing it to get stuck in a local minimum. Moreover, the stability of these losses in more complex scenarios, such as with many harmonics, transients, or inharmonicity remains an open question. The SOT loss was found to be highly sensitive to small frequency variations, such as in spectral side lobes or noise, and in our preliminary experiments, it struggled with more realistic datasets such as Nsynth \cite{engelNeuralAudioSynthesis2017}.

\section{Conclusion}
\label{sec:conclusion}

In this work, we leverage optimal transport theory to develop a training objective for audio reconstruction tasks. The proposed Spectral Optimal Transport (SOT) loss, which compares discrete spectra \textit{horizontally} (measuring energy displacement), overcomes some limitations of traditional spectral losses. We have shown that it has the potential for frequency localization and joint estimation of fundamental frequency ($f_0$) and harmonic amplitudes. Results on a synthetic dataset have demonstrated the efficacy of our method compared to the commonly employed Multi-Scale Spectral (MSS) loss.

However, our findings also reveal that the proposed loss is very sensitive to small spectral variations and might not suffice, in its current state, for accurate  $f_0$ matching in real signals. This underscores the need for a deeper understanding of the trade-offs involved.  

\clearpage

\vfill\pagebreak

\bibliographystyle{IEEEbib}
\bibliography{refs}

\end{document}